# Near-forward Raman selection rules of the phonon-polariton created by alloying in (Zn,Be)Se

H. Dicko,[1] O. Pagès,[1,a] F. Firszt,[2] K. Strzałkowski,[2] W. Paszkowicz,[3] A. Maillard,[4] C. Jobard,[4] and L. Broch[1]

[1] LCP-A2MC, Institut Jean Barriol, Université de Lorraine, France

[2] Institute of Physics, N. Copernicus University, 87-100 Toruń, Poland

[3] Institute of Physics, Polish Academy of Sciences, 02-668 Warsaw, Poland

[4] LMOPS, Université de Lorraine – Sup´elec, 2, rue Edouard Belin, 57070 Metz, France

## Abstract

The Raman selection rules of the (Zn-Se,Be-Se)-mixed phonon-polariton (PP) created by "alloying" in the three-mode [1x(-Zn-Se),2x(Be-Se)] $Zn_{1-x}Be_xSe$ system, whose dramatic S-like dispersion (~200 cm$^{-1}$) covers the large frequency gap between the Zn-Se and Be-Se spectral ranges, is studied in its wave vector ($q$) dependence by near-forward scattering. Both the collapse regime away from Γ and the reinforcement regime near Γ are addressed, using appropriate laser lines and Be contents. We find that in both regimes the considered PP, in fact a transverse mode with mixed mechanical-electrical character, obeys the same nominal Raman selection rules as its purely-mechanical (PM) variant commonly observed in the backscattering geometry. Besides, marked differences in the PP Raman lineshapes in the two regimes give a hint about how the PP-like electrical field $\vec{E}$ develops while descending the S-like dispersion towards Γ. In the reinforcement regime $\vec{E}$ is large, leading to intra-mode on top of inter-mode $\vec{E}$—mediated transfers of oscillator strength between the two Be-Se modes, that both exhibit a fine structure on account of the alloy disorder. In contrast, in the collapse regime $\vec{E}$ remains weak, as testified by the absence of intra-mode transfer. The discussion is supported by contour modeling of the multi-PP Raman lineshapes in their $q$-dependence within the linear dielectric approach.

---

[a] Author to whom correspondence should be addressed :
Email: Olivier.pages@univ-lorraine.fr

## I. Introduction

In such a polar crystal as a zincblende-type semiconductor crystal, the transverse optic (TO) phonons corresponding to vibrations of the anionic sublattice against the cationic one perpendicularly to the direction of propagation $\vec{q}$, are likely to be accompanied by an electrical field, notably a transverse one, hence identical in nature to that carried by a pure transverse electrical wave, namely a photon.[1-3] The resulting TO excitation with mixed mechanical-electrical character is referred to as a phonon-polariton, abbreviated PP hereafter. Regarding applications, the PP's are especially interesting in view of signal processing at terahertz (phonon-like) frequencies at light-like (photon-like) speed.[4]

One remarkable feature of the photon-like transverse electrical field carried by a TO mode is that it can exist only very close to the centre $\Gamma$ ($q = 0$) of the Brillouin zone. This is because the '$\omega$ vs. $q$' dispersion of a photon, governed by the speed of light in the crystal (scaled down from that in vacuum, i.e. $c$, via the relative dielectric function $\varepsilon_r$ of the crystal), is quasi vertical on the $q$ scale given by the Brillouin zone size. Far from $\Gamma$, i.e. from the natural dispersion of a photon, a transverse electrical field cannot propagate, and a TO mode then reduces to a purely-mechanical vibration. The corresponding frequency is referred to as $\omega_T$ hereafter. Note that the electrical field carried by a longitudinal optic (LO) mode does not suffer such restriction, because it is not photon-like. As such it remains non dispersive (not $q$-dependent), meaning that a LO mode vibrates at a fixed frequency near $\Gamma$, noted $\omega_L$. As is well-known $\omega_L > \omega_T$ since the spring-like restoring force of a purely-mechanical TO mode is enhanced by the Coulombian interaction resulting from the LO-like electrical field.[5]

The picture for PP's propagating in the bulk of pure zincblende compounds is well established both theoretically and experimentally.[6-8] Basically it can be casted into a model of a single (TO,LO) oscillator. A strong PP-coupling occurs when the horizontal phonon-like dispersion of a purely-mechanical TO mode away from $\Gamma$, defined by $\omega_T$, crosses the quasi vertical photon-like dispersion of a pure transverse electrical field near $\Gamma$. This gives rise to an anticrossing, resulting in two distinct PP branches. The upper one, labeled $PP^+$, is LO-like ($\omega_{PP}^+ \sim \omega_L$) near $\Gamma$ (strictly at $\Gamma$ the TO–LO degeneracy even occurs since the atom displacement can as well be viewed as perpendicular or parallel to $\vec{q}$, the latter being null in fact) and photon-like away from the phonon resonance ($\omega \gg \omega_T$); the dispersion is then dictated by $\varepsilon_\infty$, the relative dielectric constant of the crystal at 'infinite' frequency. The lower branch, noted $PP^-$, is photon-like well beneath the phonon resonance ($\omega \ll \omega_T$); the dispersion is then governed by the static relative dielectric constant $\varepsilon_s$, linked to $\varepsilon_\infty$ via the Lyddane-Sachs-Teller relation. This lower branch ultimately assimilates the purely-mechanical TO mode ($\omega_{PP}^- \sim \omega_T$) away from $\Gamma$.

Little attention has been awarded to the PP's propagating in multi-oscillator systems such as mixed crystals, the central issue in this work. Based on Maxwell's equations, Poulet produced a generic calculation of the PP dispersion in a multi-oscillator assembly already in the seventies, that was useful to set the trend.[9] Recently Bao et al. calculated the dispersion of the bulk and surface PP's of various $A_{1-x}B_xC$ zincblende mixed crystals, taken either as pseudo-infinite[10] or confined[11] (at one or two dimensions) media. In doing so they assumed a basic 2-mode [1×(A-C),1×(B-C)] behavior of the purely-mechanical TO modes from which the PP's proceed. On the experimental side, the contributions are rather sparse. We are only aware of the pioneering study of the surface PP's of the one-mode $Al_{1-x}Ga_xN$ and $Al_{1-x}In_xN$ mixed crystals with wurtzite structure done by Ng et al.[12] and by Ooi et al.,[13] respectively, using far-infrared attenuated total reflectance, and of our own recent

Raman studies of the bulk PP's of the three-mode [1×(Zn-Se),2×(Be-Se)] $Zn_{1-x}Be_xSe$[14] and [1×(Zn-Se),2×(Zn-S)] $ZnSe_{1-x}S_x$[15,16] zincblende systems.

What emerged from the above theoretical and experimental studies is that the $A_{1-x}B_xC$ "alloying" creates an intermediary $PP^{int.}$ branch with mixed (A-C,B-C) character between the lower $PP^-$ and upper $PP^+$ parent-like ones, now mostly related to the long/soft (say A-C) and short/stiff (B-C then) bonds, respectively. The $PP^{int.}$ branch differs in nature from the $PP^-$ and $PP^+$ ones in that it exhibits a characteristic '$\omega$ vs. $q$' S-like dispersion throughout the frequency gap between the B-C (upper) and A-C (lower) spectral ranges, governed by two asymptotic phonon frequencies, and not by any photon-like one. These are, away from $\Gamma$, the frequency of the BC-like purely-mechanical TO mode, and, near $\Gamma$, the frequency of the AC-like LO mode. An example of the above described PP dispersion is shown, e.g., in Ref. **15** for the particular $ZnSe_{0.68}S_{0.32}$ mixed crystal (see Fig. 1 therein). The $PP^{int.}$ mode is further interesting with respect to the $q$-dependence of its Raman intensity. The latter is large in the two asymptotic matter/phonon-like regimes (LO-like near $\Gamma$, and purely-mechanical TO-like away from $\Gamma$) and small in the intermediary light/photon-like one, corresponding roughly to the inflexion of the S-branch. In fact, the TO-like downfall away from $\Gamma$ and subsequent LO-like reinforcement near $\Gamma$ nicely come out from calculations when using our generic expression of the $q$-dependent multi-PP Raman cross section (see Ref. **16**, recalled below). The two regimes have further been observed experimentally by Raman scattering, albeit separately, i.e. in $Zn_{0.5}Be_{0.5}Se$ (see Fig. 4 of Ref. **14**) and $ZnSe_{0.68}S_{0.32}$ (see Fig. 5 of Ref. **15**), respectively.

In the case of a multi-oscillator mixed crystal, i.e. falling beyond the basic two-mode (A-C and B-C description, such as the three-oscillator [1x(Zn-Se),2x(Be-Se)] $Zn_{1-x}Be_xSe$ and [1x(Zn-Se),2x(Zn-S)] $ZnSe_{1-x}S_x$ systems, the $PP^{int.}$ branch duplicates. It becomes accompanied by a minor replica at high frequency, referred to as $PP^{min.}$ hereafter, similar in every respect to $PP^{int.}$, notably in what regards the S-like shape, the double-phonon asymptotes and also the collapse and reinforcement regimes, except that its dispersion remains confined within the relevant (Be-Se or Zn-S) doublet of purely-mechanical TO's. The picture can be naturally extrapolated beyond the three-oscillator case.

In this work our aim is to pursue the basic characterization of the main 'alloy'-related $PP^{int.}$ mode by examining its Raman selection rules both in its TO-like collapse regime and in its LO-like reinforcement one, using the same $A_{1-x}B_xC$ mixed crystal for the sake of consistency (possibly playing with the composition $x$ and with the laser excitation, though). We are only aware of two symmetry analysis of the PP Raman modes in the literature, done with the pure GaP[7] and GaN[17] compounds, and thus concerned with the ($PP^-$, $PP^+$) modes. As for mixed crystals, the $PP^{int.}$ (and also $PP^{min.}$) Raman selection rules remain unexplored so far. In our case it is a matter to elucidate whether the $PP^{int.}$ mode created by "alloying" near $\Gamma$ retains the symmetry of the purely-mechanical TO mode away from $\Gamma$ from which it proceeds, or not, and also, whether the PP symmetry is preserved when passing on from the TO-like collapse regime to the LO-like reinforcement one, or not. On the experimental side certain conditions have to be fulfilled in view to tackle such issues, concerning both the used Raman scattering geometry and also the samples, as detailed below.

Besides, we take this occasion of the planned comparison between the TO-like collapse and LO-like reinforcement regimes of the $PP^{int.}$ mode, to further investigate an intriguing feature recently evidenced with $ZnSe_{0.68}S_{0.32}$. This feature, so far unexplained, is that the $PP^{int.}$ peak drastically sharpens when it departs from the collapse regime and enters the reinforcement one (see Figs. 5 and 6 of Ref. **15**). One may well wonder whether the trend is specific to $ZnSe_{1-x}S_x$, for some reason, or persists with another mixed crystal, being thus presumably general. In this case, the next issue would be to identify its possible mechanism.

## II.    Experiment: Raman scattering geometry and samples

Generally, optical spectroscopies, such as Raman scattering, are well suited to study the PP's, because the quasi-vertical dispersion of the laser probe places, by construction, the analysis near Γ. However, the access to the PP regime is not obvious for all that. For example, in the traditional backscattering geometry (schematically operating in 'reflection'), the scattering angle $\theta$ between the wave vectors of the incident laser beam ($\vec{k}_i$) and of the scattered light ($\vec{k}_s$) is at maximum (~180°). In this case, one addresses the largest achievable phonon wavevector in a Raman scattering experiment, with magnitude of the order of a few percent of the Brillouin zone size. This falls deep into the asymptotical regime of purely-mechanical TO modes.[6] In order to access the PP regime, the phonon wave vector has to be reduced by as much as two orders of magnitude. This requires to adopt very small $\theta$ values, typically below 5°, corresponding to a near-forward scattering geometry (schematically operating in 'transmission').[6] Even in this case, generally one can hardly penetrate the PP dispersion towards Γ farther than the first half.[6]

Also, the choice of system is crucial. Out of the two samples series at hand whose PP's were successfully detected by near-forward Raman scattering so far, Zn$_{1-x}$Be$_x$Se[14] is preferred to ZnSe$_{1-x}$S$_x$[15] on account of the dramatic S-like dispersion of its $PP^{int.}$ mode, essentially due to the large frequency gap between the vibration spectral ranges of its constituent bonds (~200 cm$^{-1}$). In ZnSe$_{1-x}$S$_x$ the frequency gap is so narrow (merely ~40 cm$^{-1}$) that the $PP^{int.}$ mode enters almost straightaway its LO-like reinforcement regime without being detectable in the preceding TO-like collapse regime (when descending the S-like dispersion towards Γ). The latter, shrunk to a minimum, is somehow skipped out.[15]

Appropriate Zn$_{1-x}$Be$_x$Se samples are selected based on the intuitive 'rule of thumb' – verified theoretically (see below), that the frequency domain covered by the TO-like collapse regime of the $PP^{int.}$ mode near Γ is large when the corresponding native signal of purely-mechanical TO mode away from Γ, namely the BeSe-like one in this case, is strong, and vice versa. With this respect, large Be contents seem optimal in view to explore the Raman selection rules in the TO-like collapse regime. Reversely, small Be contents may facilitate an access to the LO-like reinforcement regime closer to Γ, due to a shrinkage of the preceding TO-like collapse regime farther from Γ. However, care should be taken that the Be content be not too small, otherwise the penetration downward the S-like dispersion towards Γ would be less, thus reducing the chance to access the LO-like reinforcement regime. This results from the wave vector conservation law $\vec{q} = \vec{k}_i - \vec{k}_s$ governing the Raman scattering. Accordingly minimal $q$ values are achieved by reducing the dispersion of the refractive index around the used laser line (refer to **Eq. 2** below). With Zn$_{1-x}$Be$_x$Se, this requires large Be contents and low energy laser lines.[18]

Three large-size Zn$_{1-x}$Be$_x$Se single crystals (cylinders, ~8 mm in diameter, ~5 mm in height) were grown from the melt by using the high-pressure Bridgman method, with small (x=0.12) and moderate (x=0.33) Be contents and with the maximum Be content (x=0.53) achievable by this technique. Owing to its low Be content, the former sample (12 at.% Be) is not relevant for the planned study of the $PP^{int.}$ mode. It is only used to provide a reference insight into the Raman selection rules of the purely-mechanical ZnSe-like TO mode (that shows up strongly) in the backscattering geometry (see **Fig. A-1a**, **Appendix Section**). The corresponding backscattering-like insight into the doublet of purely-mechanical BeSe-like TO's is likewise gained from the sample with the larger Be content (53 at.% Be, see **Fig. A-1b**). At this composition the Be-Se signal shows up most strongly and, moreover, the two Be-Se sub-mode exhibit comparable Raman intensities (see below), giving rise to an unequally well-resolved Be-Se doublet. Most of all, the latter sample (53 at.% Be), together with the remaining one (33 at.% Be), are presumably well-suited to address the TO-like collapse regime and the LO-like reinforcement one of the $PP^{int.}$ mode, respectively, as discussed above. The three ingots were prepared with parallel (within less than 1°) rear and front $(0\bar{1}1)$-oriented faces, polished to optical quality. We recall that at normal incidence onto such crystal faces the TO modes are allowed and the LO modes are forbidden, whether using the standard backscattering geometry ($\theta{\sim}180°$, for detection of the purely-mechanical TO's) or the less common near-forward ($\theta{\sim}0°$, giving access to the PP's) one.

The near-forward Raman selection rules were realized by focusing the incident laser beam at near-normal incidence onto the rear crystal face using a lens with large focal length (so as to minimize the angular spread of the incident laser beam, for better definition of $\vec{k}_i$) and by collecting the scattered light from the front crystal face in the same direction at large distance (for better definition of $\vec{k}_s$). Two half-wave plates were placed on each side of the sample with their neutral axis taken either parallel to each other or rotated by 45°, corresponding to crossed ($\perp$) and parallel ($\parallel$) polarizations of the incident laser beam ($\vec{e}_i$) and of the scattered light ($\vec{e}_s$) at the sample surface, respectively. For each configuration ($\perp$ or $\parallel$) the half-wave plates were turned as rigid bodies over 180° with a quasi constant step of 10° so as to complete full rotations of $\vec{e}_i$ and $\vec{e}_s$ (over 360°) at the sample surface. The angle between $\vec{e}_i$ and the $[\bar{1}11]$ axis of the rear and front crystal faces, taken as an arbitrary reference, is referred to as the azimuth angle hereafter, denoted $\alpha$. At each $\alpha$ value the scattered light was analyzed parallel to the entrance slit of the spectrometer, i.e. with a fixed sensitivity of the spectrometer – the maximum one in fact, so that the $\alpha$-dependent Raman intensities are directly comparable.[19]

The Raman spectra were taken with the low-energy 633.0 nm (red) and 785.0 nm (near-infrared) laser lines from a HeNe laser and from a laser diode, respectively. Note that the latter operates at the detection limit of our Raman spectrometer, blazed for the visible. No appreciable heating effect was detected in spite of the used illumination of ~50 mW, owing to large impact spot of the focused laser beam (~20 $\mu$m in diameter – recall the large focal length) resulting in a small power density at the sample surface (~10 mW/cm²).

### III.    Results and discussion

#### 1.   Theoretical insight

Prior to the experimental study, we test theoretically the presumed suitability of the selected large (53 at.%) and moderate (33 at.%) Be contents to address the TO-like collapse regime and the LO-like reinforcement one of the $PP^{int.}$ mode in $Zn_{1-x}Be_xSe$, respectively, depending on the exciting laser line. For doing so we use our generic $(x,q)$-dependent multi-PP ($p{=}3$ in total) Raman cross section (RCS, see, e.g., Ref. **16**), which we recall below for clarity,

$$RCS(\omega,q,x){\sim}Im\left\{-[\varepsilon_r(\omega,x)-q^2\times c^2\times\omega^{-2}]^{-1}\times\left[1+\sum_p C_p(x)\times L_p(\omega,x)\right]^2+\sum_p C_p^2(x)\times\right.$$
$$\left.\omega_{T,p}^2(x)\times L_p(\omega,x)\times S_p^{-1}(x)\times\varepsilon_{\infty,p}^{-1}\times\omega_{T,p}^{-2}\right\}. \qquad (1)$$

$L_p(\omega,x) = \omega_{T,p}^2(x) \times \left(\omega_{T,p}^2(x) - \omega^2 - j \times \gamma_p(x) \times \omega\right)^{-1}$ represents a damped Lorentzian resonance at the $\omega_{T,p}(x)$ frequency of the purely-mechanical $p$-type TO mode. $\varepsilon_r(\omega,x)$ is the relative dielectric function of $Zn_{1-x}Be_xSe$, expressed in its classical form, i.e. $\varepsilon_\infty(x) + \sum_p \varepsilon_{\infty,p} \times S_p(x) \times L_p(x,x_p)$, assuming a linear variation of $\varepsilon_\infty(x)$. $S_p(x)$ and $C_p(x)$ are, correspondingly, the oscillator strength and the Faust-Henry coefficient awarded to oscillator $p$. These monitor the Raman intensity of the $p$-type purely-mechanical TO mode and the splitting [in reference to $S_p(x)$] together with the Raman intensity ratio between the latter and its LO counterpart [in reference to $C_p(x)$]. Both $S_p(x)$ and $C_p(x)$ scale linearly with the fraction of oscillator $p$, i.e. as (1-x) and (x) for the unique Zn-Se oscillator and for the lower and upper Be-Se sub-oscillators taken together, respectively. The available BeSe-like oscillator strength further shares between the latter two in proportion to the fractions of the corresponding Be-like (x) and Zn-like (1-x) environments.[14] The related Faust-Henry coefficients scale accordingly. With this, the Raman intensities of the two BeSe-like purely-mechanical TO modes of $Zn_{0.47}Be_{0.53}Se$ and $Zn_{0.67}Be_{0.33}Se$, noted $TO_{Be-Se}^{Be}$ and $TO_{Be-Se}^{Zn}$ hereafter, are expected to scale roughly in the ratios 1:1 and 1:2, respectively, as actually observed in their reference Raman spectra taken in the backscattering geometry (see below). In the used notation the subscript and the superscript refer to the considered bond and to its local environment, respectively. Additional "alloy"-related input parameters are the three [1×(Zn-Se),2×(Be-Se)] $\omega_{T,p}(x)$ Raman frequencies measured in the backscattering geometry, namely (224, 447, 487) and (215, 420, 466) in cm$^{-1}$, respectively, with corresponding phonon dampings $\gamma_p(x)$ of (23, 36, 37) and (23, 38, 38), also given in cm$^{-1}$. Parent-related input parameters are also needed, in reference to the $(\varepsilon_\infty, C, \omega_T, \omega_L)$ values, corresponding to (5.75, -0.7, 205 cm$^{-1}$, 254.5 cm$^{-1}$) for ZnSe and (5.32, -0.7, 450 cm$^{-1}$, 501 cm$^{-1}$) for BeSe.[14]

By solving numerically the characteristic dispersion of a TO mode (a PP one as well as the corresponding purely-mechanical TO one), as derived from Maxwell's equations, namely $\varepsilon_r(\omega,x) = q^2 \times c^2 \times \omega^{-2}$, which leads to the divergence of the term within brackets in Eq. 1, one accesses the theoretical '$\omega$ vs. $q$' PP dispersion of the considered $Zn_{1-x}Be_xSe$ mixed crystal. The as-obtained PP dispersion curves for $Zn_{0.47}Be_{0.53}Se$ and $Zn_{0.67}Be_{0.33}Se$ are displayed in **Figs. 1a** and **1b** (thin lines), respectively, with an emphasis on the $PP^{int.}$ (and $PP^{minor}$) mode(s) as the main interest in this work. For more convenience we substitute for $q$ the dimensionless parameter $y = q \times c \times \omega_1^{-1}$, in which $\omega_1$ arbitrarily refers to the frequency of the purely-mechanical $TO_{Zn-Se}$ mode. Useful asymptotic transverse frequencies are those of the purely-mechanical $(TO_{Zn-Se}, TO_{Be-Se}^{Be}, TO_{Be-Se}^{Zn})$ modes ($q \to \infty$) and related $(LO_{Zn-Se}, LO_{Be-Se}^-, LO_{Be-Se}^+)$ ones ($q \to 0$).

An interesting question is how far one can descend the theoretical $PP^{int.}$ (and also $PP^{min.}$) 'S-like' dispersion(s) towards $\Gamma$ by near-forward Raman scattering with a given laser excitation? Such laser-dependent limits are identified by superimposing the theoretical PP dispersions onto the experimentally achievable '$\omega$ vs. $q$' ones given by the wave vector conservation law governing the Raman scattering process (quasi-linear/oblique dashed lines in **Fig. 1**), i.e.

$$q = c^{-1} \cdot \left\{ n^2(\omega_i,x)\omega_i^2 + n^2(\omega_s,x)\omega_s^2 - 2n(\omega_i,x)n(\omega_s,x)\omega_i\omega_s\cos\theta \right\}^{\frac{1}{2}}, \qquad (2)$$

taken in the limiting case of perfect forward scattering ($\theta = 0°$). We recall that by enlarging $\theta$ to a finite value one retreats away from $\Gamma$, whichever laser line or/and alloy composition are used (examples are given below). The dependence on the exciting laser line is contained in the $n(\omega_i,x)$ and $n(\omega_s,x)$ values, representing the $Zn_{1-x}Be_xSe$ refractive indexes at the frequencies of the incident laser beam ($\omega_i$) and of the scattered lights ($\omega_s$). The wavelength ($\lambda$) dependencies in the visible spectral range of the refractive indexes of our two $Zn_{1-x}Be_xSe$ crystals, measured by ellipsometry, were accurately fitted to the Cauchy dispersion formula $n(\lambda) = X + Y \times \lambda^2 \times 10^4 + Z \times \lambda^{-4} \times 10^9$, using $(X,Y,Z)$ values of $(2.3767 \mp 0.0015, 2.0873 \mp 0.0792, 3.1818 \mp 0.0916)$ for x=0.53 and of $(2.4187 \mp 0.0003, 2.2111 \mp 0.0225, 4.9053 \mp 0.0317)$ for x=0.33, where $X$ is dimensionless, $Y$ and $Z$ are constants with units of $nm^2$ and $nm^4$, respectively, and $\lambda$ is in $nm$.

The as-obtained $PP^{int.}$ (and also $PP^{min.}$) limits for $Zn_{0.47}Be_{0.63}Se$ and $Zn_{0.67}Be_{0.33}Se$ for various laser lines at hand, i.e. not only the presumably best-suited near-infrared (NIR, 785.0 nm) and red (R, 633.0 nm) ones, as discussed above, but also the more energetic green (G, 514.5 nm) and blue (B, 488.0 nm) ones from an Ar+ laser, also considered for the sake of completeness, are indicated by filled (open) symbols in **Fig. 1**. As expected (see **Sec. II**), less energetic laser lines penetrate deeper downward the $PP^{int.}$ dispersion towards Γ. Nevertheless, in either crystal the experimental insight hardly gets beyond the first half of the $PP^{int.}$ dispersion, meaning that the very close neighborhood of Γ remains out of reach.

The next question is whether such moderate penetration suffices to achieve the LO-like reinforcement regime beyond the TO-like collapse one, or not? For a direct insight we complete our theoretical insight in **Fig. 1** by adding the $RCS(\omega, \theta = 0°, x)$ PP Raman lineshapes corresponding to the above identified limits (thin lines). The calculations were done by using a variant of **Eq. 1** in which $\theta$ (taken equal to 0°) substitutes for $(q, y)$ along the univocal correspondence given by **Eq. 2**. The native purely-mechanical $TO_{Be-Se}^{Be}$ and $TO_{Be-Se}^{Zn}$ $RCS(\omega, \theta = 180°, x)$ Raman lineshapes (emphasized in grey) and related $LO_{Be-Se}^{-}$ and $LO_{Be-Se}^{+}$ $RCS(\omega, q=0, x)$ Raman lineshapes (dashed lines) are also shown, for reference purpose. In any case a minimal phonon damping $\gamma_p(x)$ of 1 cm$^{-1}$ is taken, for clarity.

While the most energetic blue and green laser lines can only address the TO-like collapse regime in its early stage, the red excitation covers the latter regime in full, down to complete $PP^{int.}$ extinction. However, the red line fails to enter frankly the LO-like reinforcement regime of our two mixed crystals (compare the theoretical Raman intensities of the $PP^{int.}$ peak and of the corresponding native $TO_{Be-Se}^{Be}$ and $TO_{Be-Se}^{Zn}$ features in **Fig. 1** – thick lines). The only successful excitation with this respect is the near-infrared (785.0 nm) one, thereby emerging as the most relevant laser line in view of probing the Raman selection rules in the LO-like reinforcement regime.

## 2. Raman selection rules in the near-forward scattering geometry

We display in **Fig. 2** the unpolarized TO-like Raman spectra of our three $Zn_{1-x}Be_xSe$ samples taken in the traditional backscattering geometry (corresponding to the theoretical peaks emphasized in grey in **Fig. 1**) using the 633.0 nm laser line, for reference purpose. We recall that in the backscattering geometry, one falls deep into the asymptotic regime of purely-mechanical ($TO_{Zn-Se}$, $TO_{Be-Se}^{Be}$, $TO_{Be-Se}^{Zn}$) modes. The corresponding frequencies, which are not dependent on the used laser line, occur near 210 cm$^{-1}$, 425 cm$^{-1}$, 500 cm$^{-1}$, with slight variations depending on the Be content – see **Sec. III-1**). The purely-mechanical TO's are flanked on their high frequency side, i.e. around 240 cm$^{-1}$, 430 cm$^{-1}$ (not observed[14]), 525 cm$^{-1}$, by their ($LO_{Zn-Se}$, $LO_{Be-Se}^{-}$, $LO_{Be-Se}^{+}$) counterparts. These LO "shadows" are theoretically forbidden, but activated as minor peaks due to a disorder-induced breaking of the wave vector conservation law, a common feature in mixed crystals. Alternative peaks/bands (notably visible at ~200 and ~325 cm$^{-1}$) reflect maxima in the one-phonon (disorder-activated) and two-phonon (allowed) density of states involving the transverse (TA) and/or longitudinal (LA) acoustical modes. Among these modes, notably the so-called X mode, that emerges roughly halfway the Zn-Se and Be-Se spectral ranges, plays an important role in the following discussion. Based on recent measurements of the $Zn_{0.67}Be_{0.33}Se$ phonon dispersion by inelastic neutron scattering, and on related *ab initio* calculations,[20] the X mode is tentatively identified as the two-phonon TA+LA band from the X zone-edge.

In anticipation of the forecoming discussion of the near-forward Raman spectra, we mention that, somewhat surprisingly, the purely-mechanical ($TO_{Zn-Se}$, $TO_{Be-Se}^{Be}$, $TO_{Be-Se}^{Zn}$) modes remain visible in the latter Raman spectra (see below), while, in principle, they should have been replaced by their PP variants. The reason is that the $Zn_{1-x}Be_xSe$ mixed crystals exhibit large optical band gaps, inherited from their parent ZnSe (2.5 eV) and BeSe (5.5 eV) compounds.[21] As such, they are transparent to the exciting (visible) laser beam, which, therefore, undergoes multi-reflection between the rear and front crystal faces. In particular, the multi-reflection magnifies the LO modes, especially at near-normal incidence ($\theta\sim0°$).[3] More important, on its way back to the rear face, the laser beam generates a spurious backscattering signal which superimposes with the nominal PP one created by the laser beam on its way forth to the front face. In fact, the spurious BeSe-like backscattering signal totally screens the $PP^{min.}$ mode which falls into the $TO_{Be-Se}^{Be} - TO_{Be-Se}^{Zn}$ band. Such problem does not arise for the $PP^{int.}$ mode, of central interest in this work, since the latter progressively shifts away from the native ($TO_{Be-Se}^{Be}$, $TO_{Be-Se}^{Zn}$) modes when $\theta$ reduces.

Our concern in the following is to identify possible $PP^{int.}$ candidates on which to realize the Raman selection rules in both the collapse and reinforcement regimes.

In Ref. [14] (see Fig. 4) the TO-like collapse regime of the $PP^{int.}$ mode of $Zn_{0.47}Be_{0.53}Se$ could be sequenced into a detailed series of spectra corresponding to various scattering angles by using the red (633.0 nm) laser line, until quasi $PP^{int.}$ extinction at ultimately small angle ($\theta\sim0°$). From the data reported there, the most suitable scattering angle for the planned near-forward Raman selection rules in the present work appears to be a small but finite one around 0.8°. At this limit, the $PP^{int.}$ mode is probed approximately halfway to its collapse regime (at $\sim400$ cm$^{-1}$), where it remains reasonably strong and at the same time well separated from both the spurious BeSe-like backscattering signal and the X band. A direct insight is given in **Fig. 2** (the truncated near-forward BeSe-like Raman signal is shifted beneath the raw backscattering-like spectrum). Such $PP^{int.}$ mode is ideal for reliable contour modeling and subsequent quantitative analysis of the $PP^{int.}$ Raman intensity depending on the azimuth angle $\alpha$ in a given polarization setup ($\perp$ or $\parallel$) in the collapse regime.

Contrary to expectations (referring to **Fig. 1a**), when using the 785.0 nm laser line, we failed to observe the $PP^{int.}$ mode of $Zn_{0.47}Be_{0.53}Se$ in its LO-like reinforcement regime. We attribute this to the X band at $\sim325$ cm$^{-1}$ being too broad. The frequency domain that the latter covers up to 350 cm$^{-1}$, and thus partly overlaps onto the spectral range in which the LO-like reinforcement regime in this mixed crystal starts (see **Fig. 1a**). Our view is that as soon as the zone-centre $PP^{int.}$ mode enters its LO-like reinforcement regime, it resonantly decomposes into two zone-edge modes with opposite wave vectors from the X band (so as to comply with the wave vector conservation law), and thus disappears.

With respect to the X mode, $Zn_{0.67}Be_{0.33}Se$ is more attractive than $Zn_{0.47}Be_{0.53}Se$. Indeed, its X band appears at a lower frequency ($\sim300$ cm$^{-1}$, see **Fig. 1b**) and, furthermore, the LO-like regime of its $PP^{int.}$ mode starts at a higher frequency ($\sim390$ cm$^{-1}$, see **Fig. 1b**). Such 'detuning' eliminates the risk of a decomposition of the native reinforced $PP^{int.}$ mode into the X band. In fact, the reinforced $PP^{int.}$ mode of $Zn_{0.67}Be_{0.33}Se$ was successfully detected with the 785.0 nm laser line. At ultimately small scattering angle ($\theta\sim0°$), we obtain a well-resolved $PP^{int.}$ feature (shifted beneath the backscattering-like Raman signal in **Fig. 2**), emerging strong and sharp at intermediary distance between the X band and the spurious purely-mechanical ($TO_{Be-Se}^{Be}$, $TO_{Be-Se}^{Zn}$) modes. The latter $PP^{int.}$ feature is a choice candidate for realizing the Raman selection rules in the reinforcement regime.

The integrated Raman intensities of the above two selected $PP^{int.}$ modes in the collapse regime of $Zn_{0.47}Be_{0.53}Se$ and in the reinforcement regime of $Zn_{0.67}Be_{0.33}Se$ were estimated via a careful contour modeling using Lorentzian/Gaussian functions (examples are given in **Fig. 2**, emphasized in light grey) in their dependence on the azimuth angle $\alpha$, in crossed ($\perp$, open symbols) and parallel ($\parallel$, filled symbols) polarizations at the same sample spot. The resulting patterns, with characteristic greek-cross-like ($\perp$) and butterfly-like ($\parallel$) shapes, are shown in **Figs. 3a** and **3b**, respectively. In these figures, the data were normalized to the area of the most intense $PP^{int.}$ peak out of the $\perp$-type $\alpha$-series, arbitrarily taken as a reference. Quasi perfect redundancy is achieved over a complete $\alpha$ −revolution (over 360°, refer to the arrows) indicating that the laser power is stable.

The experimental greek-cross-like ($\perp$) and butterfly-like ($\parallel$) patterns compare fairly well with the corresponding theoretical ones ($\perp$-thin and $\parallel$-thick lines) calculated by expressing the Raman tensors of our zincblende crystals in the relevant $\{\frac{1}{\sqrt{2}}[0\bar{1}1], \frac{1}{\sqrt{6}}[211], \frac{1}{\sqrt{3}}[\bar{1}11]\}$ basis for a near-normal incidence onto the $(0\bar{1}1)$ crystal face. We conclude that the $PP^{int.}$ mode, whether taken in its TO-like collapse regime or in its LO-like reinforcement one, obeys the same nominal Raman selection rules as observed in the traditional backscattering geometry with the native purely-mechanical TO modes (see **Fig. A1**, **Appendix Section**). This must not be surprising since $\vec{u}$, the relative displacement of the cationic and anionic species characteristic of an optic mode, as well as its accompanying electrical field $\vec{E}$, which, taken together, govern the Raman scattering due to purely-mechanical TO's ($\vec{u}$), PP's [$\vec{u}, \vec{E}(\vec{q})$] and LO's ($\vec{u}, \vec{E}$), do both consist of first-rank tensors (vectors). The Raman tensors, which reveal the modulation of the electronic susceptibility by $\vec{u}$ and/or by $\vec{E}$, thus keep the same form for all these (see, e.g., Ref. **3**, p. 380), with concomitant impact on the Raman selection rules.

### 3. Different $PP^{int.}$ Raman lineshapes in the reinforcement and collapse regimes

Though not of direct use for the above $PP^{int.}$ symmetry analysis – the main issue in this work – we cannot escape a brief discussion of two intriguing features related to the $PP^{int.}$ mode of $Zn_{0.67}Be_{0.33}Se$ we used to perform the Raman selection rules in the reinforcement regime. As apparent from **Fig. 2**, the $PP^{int.}$ mode shows up at a lower frequency (by as much as $\delta \sim 17$ cm$^{-1}$) and is more narrow (roughly by a factor three, referring to the full width at half maximum – FWHM – of the Raman line), than the corresponding theoretical $PP^{int.}$ Raman signal (thin curve) calculated via **Eq. 1** (at $\theta \sim 0°$ for the 785.0 nm laser line) on the basis of the nominal two-mode ($TO_{Be-Se}^{Be}$, $TO_{Be-Se}^{Zn}$) description of the native BeSe-like purely-mechanical TO's (thick curves). In particular, the frequency aspect is difficult to conceive since, for a given laser excitation, there is no experimental means to further reduce the $q$ value, and hence the $PP^{int.}$ frequency, once the ultimate $\theta = 0°$ scattering angle has been reached. A further intriguing feature is that the frequency and linewidth $PP^{int.}$ anomalies are absent in the collapse regime, as discussed below.

### 3a. Experimental insight

For a direct insight into how do the above frequency and linewidth anomalies develop as $q$ reduces, we display in **Fig. 4** an extended series of near-forward Raman spectra taken with $Zn_{0.67}Be_{0.33}Se$ at various scattering angles using the 785.0 nm laser line. The corresponding frequency ($\omega$) dependencies of both the FWHM (triangles) and the Raman intensity ($I_R$, integrated over the area of the peak, circles) are displayed in **Fig. 5**. Such quantities were measured from careful Lorentzian-like contour modeling (thin curves) of the $PP^{int.}$ peaks (shifted beneath the experimental curves in **Fig. 4**) obtained after subtraction of the spurious backscattering-like signal (referring to the spectrum corresponding to full extinction of the $PP^{int.}$ mode, marked by an open square in **Fig. 4**) from the raw Raman data (thick curves). Last, before common exposure in **Fig. 5**, the FWHM and $I_R$ values were normalized to the corresponding ones of the native purely-mechanical $TO_{Be-Se}^{Be}$ mode (**Fig. 2**), taken as a reference. The theoretical '$I_R$ vs. $\omega$' dependence calculated on the basis of the nominal two-mode ($TO_{Be-Se}^{Be}$, $TO_{Be-Se}^{Zn}$) description of the native BeSe-like purely-mechanical TO's is added for comparison, down to the lowest accessible limit with the laser line used ($\theta = 0°$, 785 nm).

In principle, for each reported ($I_R$,$\omega$) couple of experimental data, the scattering angle $\theta$ can be inferred either from the Raman frequency ($\omega$) or the Raman intensity ($I_R$), or, ideally, from both. This is done by computing **Eq. 1** after substitution of $\theta$ for ($q$,$y$) along **Eq. 2**, and by adjusting $\theta$ until perfect agreement is achieved between the experimental and theoretical $PP^{int.}$ signals for the selected marker(s) (in reference to $I_R$ or/and $\omega$). However, in the present case '$\omega$' is manifestly not a reliable marker, in that, in **Fig. 5**, most of the measured Raman frequencies are massively red-shifted away from the theoretically accessible frequency domain by as much as ~17 cm$^{-1}$. Accordingly the reported $\theta$ values in **Fig. 5** were derived from a selective insight into $I_R$. Remarkably, the massive red shift is accompanied by a significant reduction in the FWHM with respect to the purely-mechanical TO's, by factor of four at ultimately small scattering angle.

In fact, the only 'decent' $PP^{int.}$ feature with respect to the theoretical '$I_R$ vs. $\omega$' two-mode (Be-Se) prediction is the first one of the series, taken at large scattering angle. Interestingly, its $I_R$ and $\omega$ values consistently indicate that it falls at the limit of the collapse regime, very close to the borderline with the reinforcement regime (identifying with the minimum of the theoretical '$I_R$ vs. $\omega$' curve). The latter $PP^{int.}$ peak is further interesting in that its linewidth matches that of its native purely-mechanical $TO_{Be-Se}^{Be}$ and $TO_{Be-Se}^{Zn}$ modes within less than ten percent (~3 cm$^{-1}$). Altogether, such data suggest that the frequency and linewidth anomalies observed with all the remaining $PP^{int.}$ modes of the series, taken at smaller scattering angles and thus coming under the reinforcement regime, are absent in the collapse regime.

However, only one $PP^{int.}$ peak of $Zn_{0.67}Be_{0.33}Se$, moreover a small one close to extinction, is not sufficient to formulate a statement valid throughout the entire collapse regime. For a fully reliable insight into the collapse regime we turn to $Zn_{0.47}Be_{0.53}Se$, in relation to the abundant data earlier published in Ref. **14** (Fig. 4 therein). Remarkably the theoretical $PP^{int.}$ Raman lineshapes calculated via **Eq. 1** on the basis of the bimodal pattern of the native purely-mechanical $TO_{Be-Se}^{Be}$ and $TO_{Be-Se}^{Zn}$ modes (thin lines in Fig. 4 of Ref. **14**) reproduce fairly well the experimental $PP^{int.}$ features at any scattering angle, in what regards both the Raman frequency ($\omega$) and intensity ($I_R$). The used implicit approximation that the FWHM of the $PP^{int.}$ mode remains identical to that of the native purely-mechanical TO's throughout the entire collapse regime is thus a valid one. For a direct insight, compare, e.g., the FWHM's of the $PP^{int.}$ (light grey), $TO_{Be-Se}^{Be}$ and $TO_{Be-Se}^{Zn}$ (dark grey) Raman lines of $Zn_{0.47}Be_{0.53}Se$ in **Fig. 3a**, obtained after contour modeling using Lorentzian/Gaussian functions. Altogether this is consistent with the $Zn_{0.67}Be_{0.33}Se$ insight in this work.

Summarizing, the available $Zn_{0.67}Be_{0.33}Se$ and $Zn_{0.47}Be_{0.53}Se$ Raman data reveal that the apparent frequency and linewidth anomalies of the $PP^{int.}$ mode come along, and appear only in the reinforcement regime, not in the collapse regime.

### 3b. The $PP^{int.}$ mode in its reinforcement regime: striking analogies with the $LO^+$ mode

Remarkably, the frequency and linewidth anomalies observed in the transverse symmetry with the $PP^{int.}$ mode of $Zn_{0.67}Be_{0.33}Se$ in its reinforcement regime are echoed in the longitudinal symmetry with the $LO^+_{Be-Se}$ mode of both the $Zn_{0.67}Be_{0.33}Se$ and $Zn_{0.47}Be_{0.53}Se$ mixed crystals. Moreover the magnitudes of both anomalies are comparable in both symmetries (refer to the frequency- and linewidth-related double arrows in **Fig. 2**). This cannot be merely fortuitous, and points towards a common cause. Only, the LO shift occurs towards high frequency (blue-shift), and not towards low frequency (red-shift) as for the $PP^{int.}$ mode.

Such anomalies of the $LO^+_{Be-Se}$ signal of $Zn_{1-x}Be_xSe$ have been studied in detail one decade ago.[23] They were attributed to a discrete (and not continuous) fine structure behind the native $TO^{Be}_{Be-Se} - TO^{Zn}_{Be-Se}$ doublet of purely-mechanical TO's. The fine structure in question results from inherent fluctuations in the local composition of a disordered mixed crystal, leading *in fine* to slight variations of the Be-Se vibration frequencies throughout the crystal. Independent support arises from recent *ab initio* calculation of the $\Gamma$-projected one-phonon density of states related to the Be atoms in a large $Zn_{64}Be_{32}Se_{96}$ disordered supercell (33 at.% Be). The latter appears to be extremely broad, expanding over ~50 cm$^{-1}$, and clearly multimodal in character (in contrast the Zn counterpart hardly covers ~10 cm$^{-1}$, and apparently consists of a unique feature - see Fig. 4 of Ref. **20**). Owing to their purely-mechanical character, the individual sub-oscillators from the fine structure do hardly couple and thus contribute on equal footing to the overall signal, that is the reason why the two main $TO^{Be}_{Be-Se}$ and $TO^{Zn}_{Be-Se}$ Raman features detected in the backscattering geometry are so broad ($\gamma_p$~37 cm$^{-1}$, see **Sec. III-1**).

Now, the situation changes in the LO symmetry. The sub-oscillators from the fine structure are then accompanied by a common macroscopic longitudinal electrical field $\vec{E}_L$ (a constant one, see **Sec. I**) via which they strongly couple. Generally, the effect of the $\vec{E}_L$-coupling is to channel the BeSe-like oscillator strength shared among various sub-oscillators of the fine structure towards a unique LO sub-mode emerging as a 'giant' one. The as-favored sub-LO mode is shifted to high frequency with respect to the remaining sub-LO's, now deprived of oscillator strength and thus repelled as minor features close to their native purely-mechanical sub-TO's. Moreover, as the 'giant' LO mode basically consists of a unique oscillator, it is sharper (referring to the FWHM) than its native purely-mechanical $TO^{Be}_{Be-Se}$ and $TO^{Zn}_{Be-Se}$ modes, which both exhibit a fine-structure (compare the FWHM of the corresponding Lorentzian/Gaussian-like profiles shown for $Zn_{0.67}Be_{0.33}Se$ in **Fig. 2**).

In principle, the above $LO^+_{Be-Se}$ picture transposes as such to the corresponding $PP^{int.}$ mode in the transverse symmetry, also equipped with a macroscopic electrical field, however, a $\vec{q}$-dependent one in this case (**Sec. I**), denoted $\vec{E}_T(q)$. One basic difference nevertheless is that the $\vec{E}_T(q)$-coupling tends to channel the oscillator strength towards low frequency, oppositely to the $\vec{E}_L$-coupling, as studied in detail in Ref. **16**.

In fact, the analogy between the $LO^+_{Be-Se}$ and $PP^{int.}$ modes is valid in the reinforcement regime of the latter − in what regards both the frequency and linewidth anomalies (refer to the Lorentzian/Gaussian-like line contour modeling displayed for the BeSe-like signal of $Zn_{0.47}Be_{0.53}Se$ in **Fig. 2**), but not in the collapse regime. Indeed, in the latter regime the FWHM of the $PP^{int.}$ mode remains identical to that of its native purely-mechanical $TO^{Be}_{Be-Se}$ and $TO^{Zn}_{Be-Se}$ modes (see **Sec. 3a**, and also **Fig. 2** − $Zn_{0.47}Be_{0.53}Se$). This indicates, in particular, the absence of any intra-mode $\vec{E}_T(q)$-mediated transfer of oscillator strength. Yet, an actual fine structure exists behind the purely-mechanical $TO^{Be}_{Be-Se}$ and $TO^{Zn}_{Be-Se}$ modes, as ascertained by the frequency and linewidth anomalies of the $LO^+_{Be-Se}$ Raman signal (see the double arrows to the LO mode of $Zn_{0.47}Be_{0.53}Se$ in **Fig. 2**).

We are thus led to conclude that the $\vec{E}_T(q)$-coupling is $\vec{q}-$dependent. Somehow this conforms to intuition since $\vec{E}_T$, the vector of coupling, is itself $\vec{q}-$dependent (in contrast with $\vec{E}_L$). In a nutshell, the coupling is strong in the reinforcement regime and inoperative in the collapse regime. A weak coupling in the collapse regime is not surprising since, in the latter regime, $\vec{E}_T(\vec{q})$ progressively departs from zero (referring to the purely-mechanical TO modes away from $\Gamma$). The reason why the $\vec{E}_T(\vec{q})$-coupling becomes operative precisely when entering the reinforcement regime is not clear yet.

### 3c. Crude theoretical insight

We must admit that we have no means to infer neither the number nor the frequencies of the sub-oscillators constituting the presumed fine structure behind the two BeSe-like purely mechanical ($TO^{Be}_{Be-Se}$, $TO^{Zn}_{Be-Se}$) modes of $Zn_{0.67}Be_{0.33}Se$. The amount of Be-Se oscillator strength awarded per sub-oscillator is likewise not accessible. For a crude quantitative insight, we artificially substitute for the apparent bimodal description ($TO^{Be}_{Be-Se}$, $TO^{Zn}_{Be-Se}$) of the purely-mechanical BeSe-like signal, an equivalent multi-mode description arbitrarily limited to ten sub-oscillators (plain line), in which the oscillator strength carried by each sub-mode of the doublet is shared among five equally-spaced (by ~14 cm$^{-1}$) sub-oscillators (awarding 32% of the oscillator strength to the central oscillator, 22% to each of its side oscillators, and 12% to each of the remaining two peripheral oscillators). A minimal phonon damping (1 cm$^{-1}$) is taken, for clarity. The crude two-oscillator (thick lines) and refined ten-oscillator (thin lines) $PP^{int.}$, purely-mechanical TO and LO theoretical $Zn_{0.67}Be_{0.33}Se$ Raman lineshapes calculated by implementing **Eq. 1** in full at the relevant $q$ value (by assuming the normal incidence, i.e. $\theta = 0°$, for the used 785.0 nm laser line), in the $q \gg \frac{\omega}{c}$ approximation and strictly at $q = 0$ (recall the TO-LO degeneracy at $\Gamma$), respectively, are displayed in **Fig. 4**. As expected, the ten-oscillator description eventually leads to the emergence of sharp $LO^+$ and $PP^{int.}$ Raman peaks, respectively blue- and red-shifted with respect to their broad two-oscillator counterparts (refer to the double-arrows in **Figs. 2** and **4**).

In brief, the observed frequency and linewidth anomalies of the polar ($\vec{E}-$equipped) $LO^+_{Be-Se}$ mode and $PP^{int.}$ mode (in its reinforcement regime) of $Zn_{0.67}Be_{0.33}Se$ find a natural explanation by assuming that their native broad purely-mechanical $TO^{Be}_{Be-Se}$ and $TO^{Zn}_{Be-Se}$ modes are finely structured due to the alloy disorder. In this case the $\vec{E}$-coupling between the individual sub-modes gives rise to an intra-mode $\vec{E}$-mediated transfer of oscillator strength on top of the classical inter-mode ($TO^{Be}_{Be-Se} \rightarrow TO^{Zn}_{Be-Se}$) one. Altogether, this results in both a drastic sharpening and an emphasized shift, towards low ($PP^{int.}$) or high ($LO^{+}_{Be-Se}$) frequency, with respect to the native purely-mechanical $TO^{Be}_{Be-Se}$ and $TO^{Zn}_{Be-Se}$ modes.

Technically, we have no flexibility to explore the effect of a weak $\vec{E}$-mediated coupling between the individual sub-oscillators, in reference to the $PP^{int.}$ mode in its collapse regime. In our simulations, the $\vec{E}$-coupling is either present (incorporated via a Coulombian force in the ionic equation of motion, as required for the $LO^{+}$ and $PP^{int.}$ modes), or absent (as relevant for the description of the purely-mechanical TO's, the Coulombian interaction being then omitted in the force assessment), with no possible modulation of $\vec{E}$.

For the sake of completeness, we report in the **Appendix Section (Sec. A-2)** a discussion of remarkable differences and analogies between the frequency and linewidth anomalies of the $\vec{E}$-equipped $LO^{+}$ and $PP^{int.}$ modes from the two ZnSe-based $Zn_{0.67}Be_{0.32}Se$ (this work) and $ZnSe_{0.68}S_{0.32}$ (Refs. **24** and **15**) mixed crystals, characterized by similar contents of their short (Be-Se, Zn-S) bond species.

## IV.     Conclusion

We perform the near-forward Raman selection rules of the $PP^{int.}$ mode created by "alloying" in the three-mode [1x(Zn-Se),2x(Be-Se)] $Zn_{1-x}Be_xSe$ system, using parallel and crossed polarizations. We find that the $PP^{int.}$ mode keeps the nominal TO symmetry as apparent in the standard backscattering geometry throughout the whole of its characteristic S-like dispersion, i.e. not only in its TO-like collapse regime away from Γ but also in its LO-like reinforcement regime near Γ. Optimal Raman insight into a given PP regime is achieved by adjusting the Be content, the laser excitation and the PP wave vector (by playing with the scattering angle $\theta$), supported by contour modeling of the PP Raman lineshapes in their ($x,q,\theta$) dependencies within the linear dielectric approach.

Remarkably, both optic vibrations that are equipped with an electrical field $\vec{E}$, namely the transverse $PP^{int.}$ one as well as the longitudinal $LO^{+}$ one, are significantly shifted (in opposite directions) and sharpened with respect to the corresponding theoretical Raman signals issued from the nominal two-mode description of the native purely-mechanical BeSe-like TO's. This reveals a disorder-induced fine structure behind the latter, leading, on top of the "classical" inter-mode $\vec{E}$-mediated transfer of oscillator strength, to an intra-mode one. With this, a unique sub-oscillator is favored out of the PP and LO series and pronounced as a giant one, both sharpened and overshifted with respect to the rest of the series. As far as the $PP^{int.}$ mode is concerned, the overshift and drastic sharpening are detected only in the reinforcement regime, the sign that $\vec{E}$ is fully developed at this limit, and not in the collapse regime, where $\vec{E}$ is only progressively departing from zero and remains weak/inoperative, notably in what regards the intra-mode transfer of oscillator strength. Altogether this provides an insight into how the photon-like transverse electrical field $\vec{E}$ of a TO mode develops when departing from the purely-mechanical ($\vec{E}=\vec{0}$) asymptotic regime away from Γ, and entering the PP regime ($\vec{E}\neq\vec{0}$) near Γ, including the collapse ($\vec{E}$ weak) and reinforcement ($\vec{E}$ strong) sub-regimes.

**Acknowledgements**

We would like to thank P. Franchetti and J.-P. Decruppe for technical assistance in the Raman measurements, and A. V. Postnikov for useful discussions and careful reading of the manuscript.

**Appendix Section**

In this **Sec.** we cover two distinct issues. In **Sec. A-1** we provide a reference insight into the Zn$_{1-x}$Be$_x$Se Raman selection rules in the traditional backscattering geometry, in relation to the purely-mechanical TO modes. In **Sec. A-2** we discuss to which extent the frequency and linewidth anomalies of the $LO^+$ and $PP^{int.}$ (in the reinforcement regime) modes related to the short bond of Zn$_{0.67}$Be$_{0.32}$Se resemble and/or differ from those of ZnSe$_{0.68}$S$_{0.32}$, earlier studied in Ref. **15**. The fractions of the short bond species (Be-Se or Zn-S) are nearly the same in the two mixed crystals, so that the related Raman signals can be directly compared.

### A1. Zn$_{1-x}$Be$_x$Se: Symmetry analysis of the purely-mechanical TO's in the backscattering Raman geometry

We report in **Fig. A1a** and **A1b** the Raman selection rules performed in the backscattering geometry with the purely-mechanical $TO_{Zn-Se}$ mode and with the like $(TO_{Be-Se}^{Be}, TO_{Be-Se}^{Zn})$ doublet (b) of Zn$_{0.88}$Be$_{0.12}$Se and Zn$_{0.47}$Be$_{0.53}$Se, respectively, using the 633.0 nm laser line. The latter crystals correspond to large Zn content and maximum Be content, respectively, and are thus well suited for such reference Zn-Se and Be-Se insights. The Raman intensities were measured in their angular ($\alpha$, see main text) dependence in a given polarization setup ($\perp$ or $\parallel$) via a careful contour modeling using Lorentzian/Gaussian functions (refer to the dark grey features in **Fig. 2**). The Be-Se contour modeling was performed directly on the raw spectra owing to the naturally well-defined baseline in this spectral range. In contrast, the $TO_{Zn-Se}$ mode is contaminated by numerous disorder-induced features, so that the subtraction of a tentative baseline (refer to the dotted curve in **Fig. 2**) was required before Lorentzian-like adjustment.

The experimental data basically conform the theoretical predictions (plain curves, see main text), in both the $\perp$ (greek-cross-like, theory – thin lines, exp. – open symbols) and $\parallel$ (butterfly-like, theory – thick lines, exp. – plain symbols) polarization setups. In fact, the agreement is nearly perfect in the case of Be-Se, with $TO_{Be-Se}^{Be}$ (circles) as well as with $TO_{Be-Se}^{Zn}$ (squares). Similar butterfly- and greek-cross-like patterns are also observed with $TO_{Zn-Se}$, but the agreement between theory and experiment is not so good, especially when the $TO_{Zn-Se}$ mode is close to extinction. This is due to the above mentioned contamination problem, which prevents a reliable determination of the Raman intensity at critical $\alpha$ angles.

The as-obtained Raman selection rules for the purely-mechanical $TO_{Zn-Se}$, $TO_{Be-Se}^{Be}$ and $TO_{Be-Se}^{Zn}$ modes ($\vec{E} = \vec{0}$) of Zn$_{1-x}$Be$_x$Se achieved in the backscattering geometry constitute natural references for the given discussion, in the main part of the manuscript (**Sec. III-2**), of the like $PP^{int.}$ ($\vec{E}$ −equipped) Raman selection rules done in the collapse and reinforcement regimes by near-forward scattering.

### A2. Zn$_{1-x}$Be$_x$Se vs. ZnS$_{1-x}$S$_x$ : comparative insight into the $PP^{int.}$ and $LO^+$ anomalies

Interestingly, the $PP^{int.}$ mode of the ZnSe$_{0.68}$S$_{0.32}$ mixed crystal exhibits a significant reduction in FWHM (referred to as the "linewidth anomaly" in the main text) when entering the reinforcement regime, as observed with Zn$_{0.67}$Be$_{0.33}$Se. However, the overshift with respect to the two-mode prediction (referred to as the "frequency anomaly" in the main text) is absent in the case of ZnSe$_{0.68}$S$_{0.32}$, in contrast to Zn$_{0.67}$Be$_{0.33}$Se. In fact, the experimental '$I_R$ vs. $\omega$' ZnSe$_{0.68}$S$_{0.32}$ data fairly match the theoretical prediction done on the basis a mere bimodal behavior of the purely-mechanical ($TO_{Zn-S}^{S}$, $TO_{Zn-S}^{Se}$) doublet, down to ultimately small scattering angle (falling into the reinforcement regime, as with $Zn_{0.67}Be_{0.32}Se$). For a direct insight we reproduce in **Fig. A2** the original '$I_R$ vs. $\omega$' plot earlier obtained with $ZnSe_{0.68}S_{0.32}$ (taken from Fig. 6 of Ref. **15**), now completed with the FWHM data.

The comparison between $Zn_{0.67}Be_{0.33}Se$ and $Zn_{0.47}Be_{0.53}Se$ can be pursued in the longitudinal symmetry with the $LO^+$ mode. As apparent in **Fig. A3**, the $LO^+$ Raman peak of $ZnSe_{0.68}S_{0.32}$ is much more narrow than the native purely-mechanical $TO_{Zn-S}^{S}$ and $TO_{Zn-S}^{Se}$ modes (see, e.g. Fig. 7 of Ref. **16**), meaning that it exhibits the linewidth anomaly, as its $PP^{int.}$ (in the reinforcement regime) counterpart. As for the frequency anomaly of $LO^+$, it is difficult to decide because it is not clear yet which oscillator strength should be used for pure ZnS when considering the $ZnSe_{1-x}S_x$ mixed crystal, as pointed out by Vinogradov *et al.* in their careful study of this mixed crystal by far-infrared reflectivity.[24] We only observe that the theoretical Zn-Se and Zn-S LO frequencies derived within the nominal three-mode [$TO_{Zn-Se}$, $TO_{Zn-S}^{S}$, $TO_{Zn-S}^{Se}$] description of the purely-mechanical TO's (**Fig. A3**, plain curves) using the nominal oscillator strengths of the pure ZnSe (2.92) and ZnS (3.00) compounds,[16] do match the experimental one (within 1 cm⁻¹) – as expected, provided the ZnSe-like LO mode is decoupled from the BeSe-like one. When calculating the LO Raman lineshape due to a given bond via **Eq. 1**, the decoupling is achieved technically by removing the oscillators due to the other bond (dashed and dotted curves) from the relative dielectric function of the crystal, and leaving the rest unchanged. A possible justification for the decoupling between the Zn-Se and Zn-S LO signals is a pining of the former signal by the numerous disorder-activated features around, as abundantly discussed in Ref. **16**.

Summarizing, the $LO^+$ and $PP^{int.}$ modes of $ZnSe_{0.68}S_{0.32}$ are consistent in what regards the presence of the linewidth anomaly along with the absence of the frequency anomaly. For comparison, in $Zn_{0.67}Be_{0.33}Se$ both anomalies occur for both modes. This reinforces the match between the linewidth and frequency anomalies of the polar ($\vec{E}$-equipped) $LO^+$ and $PP^{int.}$ (in the reinforcement regime) modes as a general feature in mixed crystals, at least in the ZnSe-based ones.

Now, an interesting question is why the linewidth anomalies of the $PP^{int.}$ (taken in the reinforcement regime) and $LO^+$ modes are not accompanied by any frequency anomaly in the case of $ZnSe_{0.68}S_{0.32}$, whereas the two anomalies come along in the case of $Zn_{0.67}Be_{0.33}Se$? To our view this relates to a basic difference in the disorder-induced fine structures behind the purely-mechanical TO's of the two mixed crystals. $Zn_{1-x}Be_xSe$ exhibits a unusually large contrast in bond length, roughly double that of $ZnSe_{1-x}S_x$ (~9% against ~4.5%, see Ref. **16**). One may thus expect larger variations in the local bond distortions of the Be-Se bonds throughout $Zn_{0.67}Be_{0.33}Se$ than those occurring for Zn-S in $ZnSe_{0.68}S_{0.32}$. Consequently, the frequencies of the individual modes forming the fine structure of the Be-Se and Zn-S doublets of purely-mechanical TO's are expected to be well separated for the former and rather close for the latter. In fact, the absence of any frequency anomaly points towards a quasi continuous fine structure for $ZnSe_{0.68}S_{0.32}$. In this case we have shown in recent work dedicated to the $LO^+$ mode of $Zn_{1-x}Be_xSe$[23] that only the linewidth anomaly occurs, and not the frequency one, consistently with experimental observations with $ZnSe_{0.68}S_{0.32}$.

It emerges that the existence/absence of a frequency anomaly for the polar ($\vec{E}$-equipped) $LO^+$ and $PP^{int.}$ (in the reinforcement regime) modes of $Zn_{0.67}Be_{0.32}Se/ZnSe_{0.68}S_{0.32}$ is subject to the discrete/continuous character of the disorder-induced fine structure behind the native purely-mechanical TO's ($\vec{E} = \vec{0}$). In contrast, the reduction in FWHM of the latter modes occurs in any case, i.e. whether the fine structuring is discrete or continuous.

**Figure captions**

**Fig. 1 :** Theoretical $Zn_{0.47}Be_{0.53}Se$ (a) and $Zn_{0.67}Be_{0.33}Se$ (b) '$\omega$ vs. $y$' PP dispersion curves (thin lines). The theoretical PP Raman lineshapes achievable experimentally in the perfect forward scattering geometry ($\theta = 0°$) by using the blue (B, 488.0 nm), green (G, 514.5 nm), red (R, 633.0 nm) and near-infrared (NIR, 785.0 nm) laser lines, calculated via **Eq. (1)** by taking a small phonon damping (1 cm$^{-1}$), are shown (thin curves), together with the corresponding purely-mechanical TO (emphasized in grey) and LO (dashed curves) ones. The relevant $(\omega, q)$ PP values for each laser line (symbols) are identified by coincidence (symbols) of the theoretical PP dispersion with the experimental '$\omega$ vs. $y$' dispersion (dotted lines) dictated by the wavevector conservation law, referring to Eq. (2). The dependence of the $PP^{int.}$ Raman intensity on the Raman frequency $\omega$ is indicated (dashed-dotted lines).

**Fig. 2:** Reference unpolarized Raman spectra taken in the traditional backscattering geometry ($\theta \sim 180°$, TO-allowed, LO-forbidden) with the $(0\bar{1}1)$-oriented $Zn_{0.88}Be_{0.12}Se$ (bottom), $Zn_{0.67}Be_{0.33}Se$ (medium) and $Zn_{0.47}Be_{0.53}Se$ (top) samples, and corresponding near-forward Raman spectra ($\theta \sim 0°$, truncated-thick curves shifted beneath the former spectra). The TO modes used to perform the Raman selection rules in the PP (light grey) and purely-mechanical (dark grey) regimes are emphasized. The $PP^{int.}$ mode of $Zn_{0.67}Be_{0.33}Se$ obtained after subtraction of the spurious backscattering signal is marked by a filled square (refer to **Fig. 4**). The theoretical BeSe-like purely-mechanical TO's (thick lines), PP's (thin lines) and LO's (dashed lines) based on a two-mode description are shown, for comparison.

**Fig. 3:** Raman selection rules in their angular ($\alpha$, see main text) dependence performed in the near-forward scattering geometry in crossed ($\perp$, Greekcross-like, open symbols) and parallel ($\parallel$, Butterfly-like, filled symbols) configurations with the $PP^{int.}$ modes (distinguished in dark grey in **Fig. 3**) of $Zn_{0.47}Be_{0.53}Se$ (a,) and $Zn_{0.67}Be_{0.33}Se$ (b), characteristic of the collapse and reinforcement regimes, respectively.

**Fig. 4:** Backward (top) and $\theta$-dependent near-forward $Zn_{0.67}Be_{0.33}Se$ Raman spectra. The corresponding $PP^{int.}$ (thin lines, bottom), purely-mechanical TO (thick lines, top) and LO (thin lines, top) theoretical Raman lineshapes, obtained on the basis of the bimodal (upper curves) and multi-mode (lower curves) descriptions of the purely-mechanical BeSe-like TO modes, are added for comparison. The experimental $PP^{int.}$ Raman signals obtained after subtraction of the near-forward Raman spectrum corresponding to total extinction of the $PP^{int.}$ mode (marked by an open square), are shifted beneath the raw data, for a clear insight..

**Fig. 5:** Frequency ($\omega$) dependencies of the FWHM (triangles) and Raman intensity ($I_R$, circles) of the $PP^{int.}$ mode of $Zn_{0.67}Be_{0.33}Se$. The theoretical '$I_R$ vs. $\omega$' curve based on a two-mode description of the Be-Se signal is added (thin line), for comparison. The scattering angle $\theta$, derived from a selective insight into $I_R$, is indicated in each case. The errors bars refer to the used procedure for contour modeling based on Lorentzian or Gaussian functions.

**Fig. A1:** Raman selection rules in their angular ($\alpha$, see main text) dependence performed in the backscattering geometry in crossed ($\perp$, Greekcross-like, open symbols) and parallel ($\parallel$, Butterfly-like, filled symbols) configurations with the purely mechanical $TO_{Zn-Se}$ mode of $Zn_{0.88}Be_{0.12}Se$ (a) and $TO_{Be-Se}^{Be}$ (circles) and $TO_{Be-Se}^{Zn}$ (diamonds) modes of $Zn_{0.47}Be_{0.53}Se$ (b), as emphasized in **Fig. 2**.

**Fig. A2:** Frequency ($\omega$) dependencies of the FWHM (triangles) and Raman intensity ($I_R$, circles) of the $PP^{int.}$ mode of $ZnSe_{0.68}S_{0.32}$. The theoretical '$I_R$ vs. $\omega$' curve based on a two-mode descriptioin of the Zn-S signal is added (thin line), for comparison. The scattering angle $\theta$, derived from combined insights into the Raman frequency ($\omega$) and the Raman intensity ($I_R$), is indicated in each case. The errors bars refer to the used procedure for contour modeling based on Lorentzian or Gaussian functions.

**Fig. A3:** representative TO-like Raman spectrum of $ZnSe_{0.68}S_{0.32}$ taken in the backscattering geometry. The corresponding theoretical three-mode $[TO_{Zn-Se}, TO_{Zn-S}^S, TO_{Zn-S}^{Se}]$ description of the purely-mechanical TO's (plain curves) is superimposed, for comparison, together with the related ZnSe- (dashed curve) and BeSe-like (dotted curve) LO Raman lineshapes. The latter were calculated by assuming a disorder-induced decoupling between the Zn-Se and Be-Se signals.

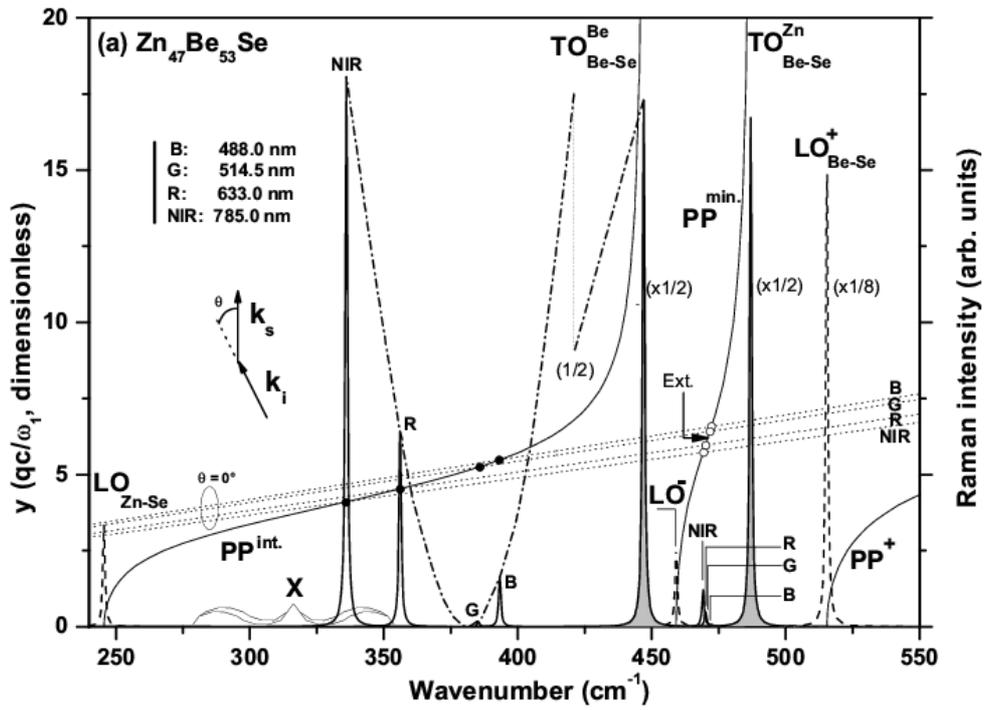

Fig. 1a

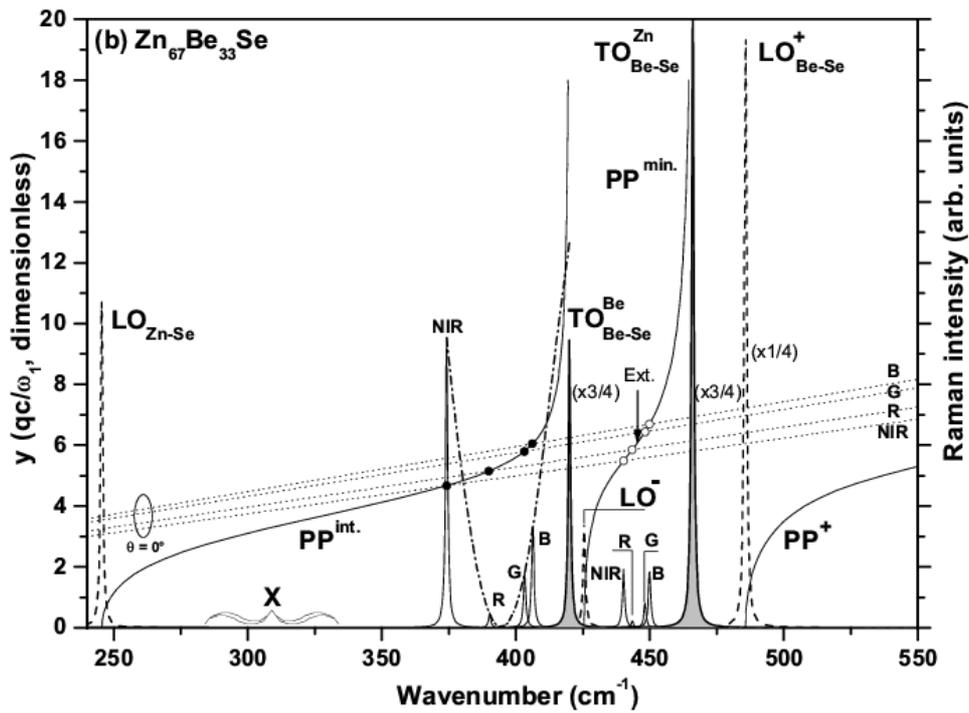

Fig. 1b

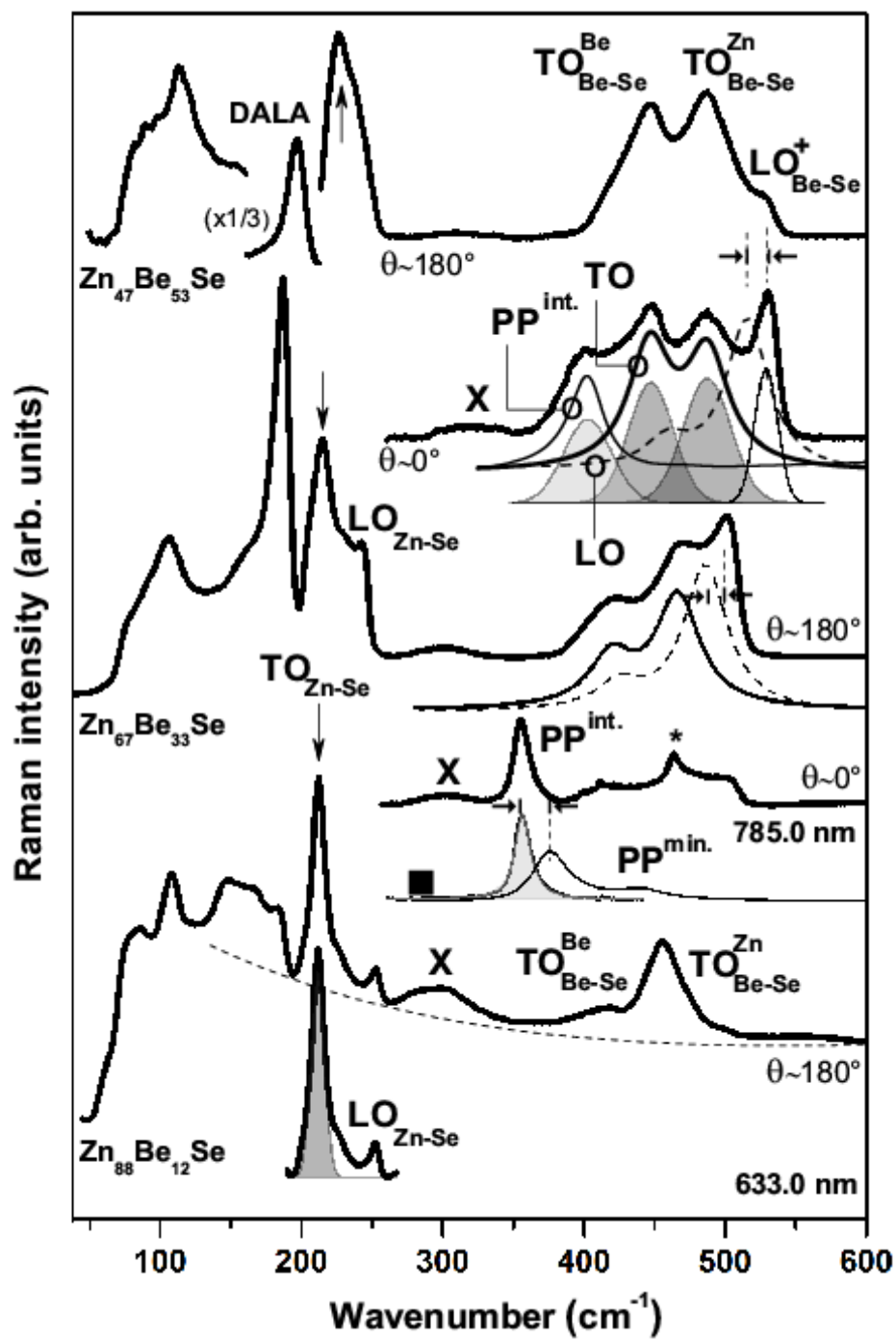

Fig. 2

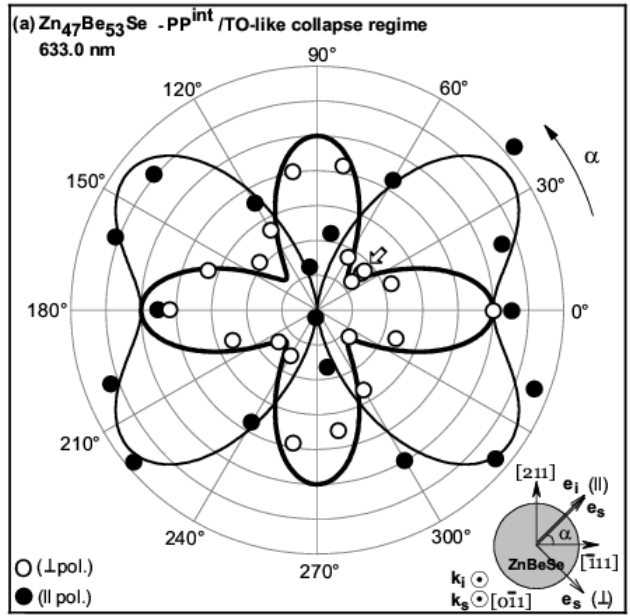

Fig. 3a

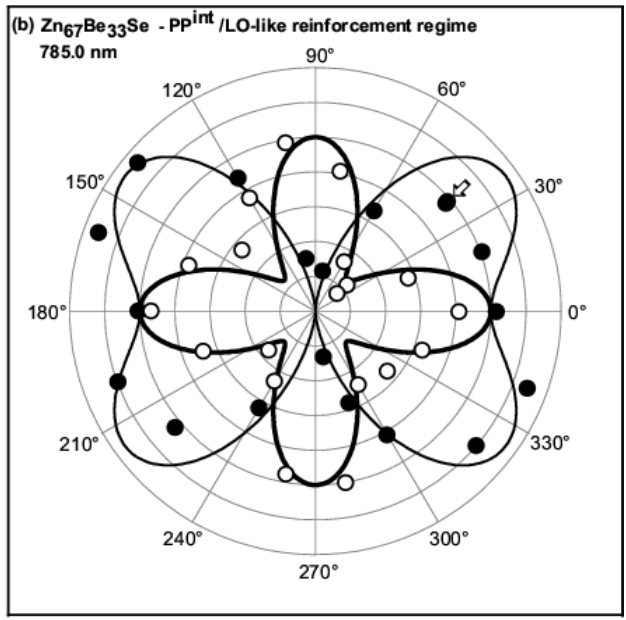

Fig. 3b

Fig. 4

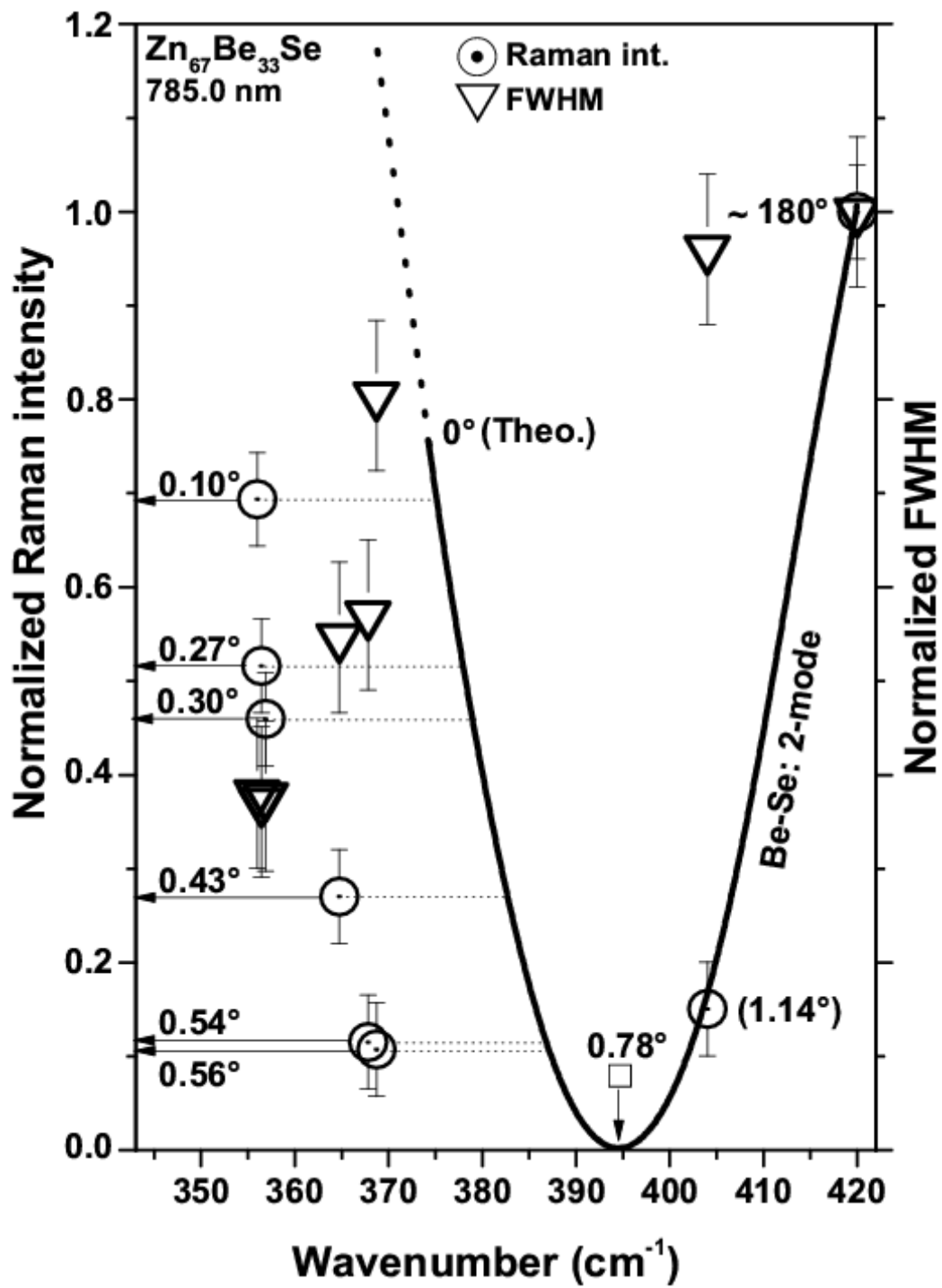

Fig. 5

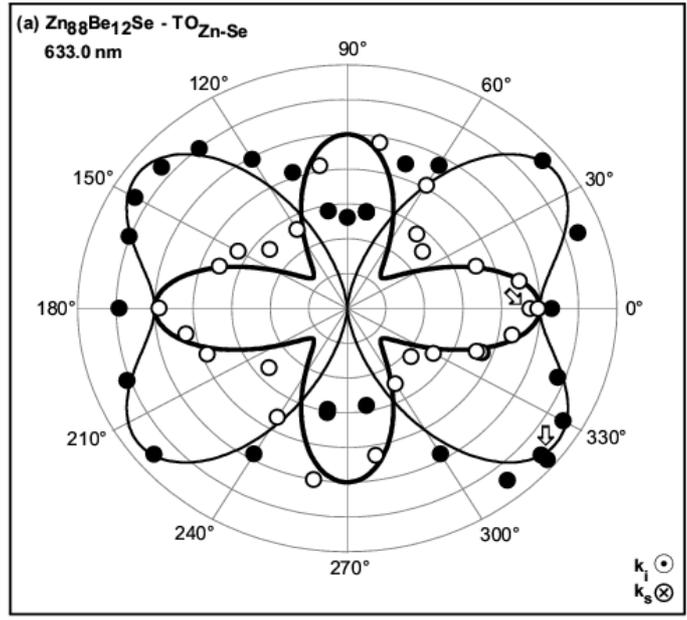

**Fig. A1a**

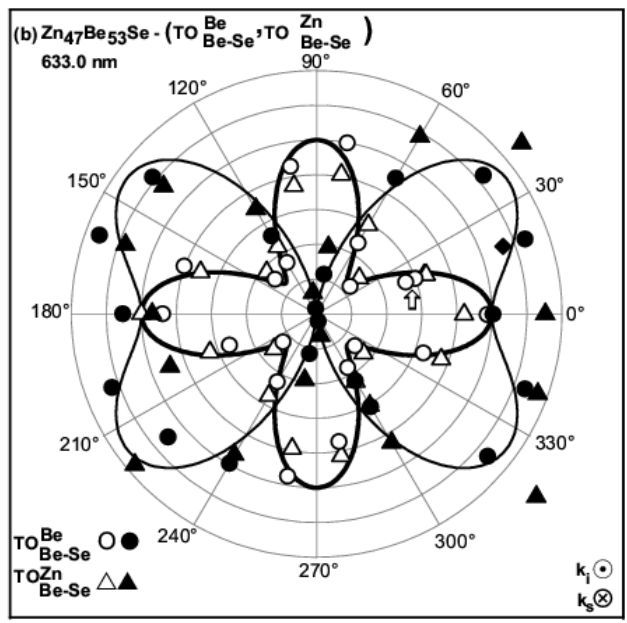

**Fig. A1b**

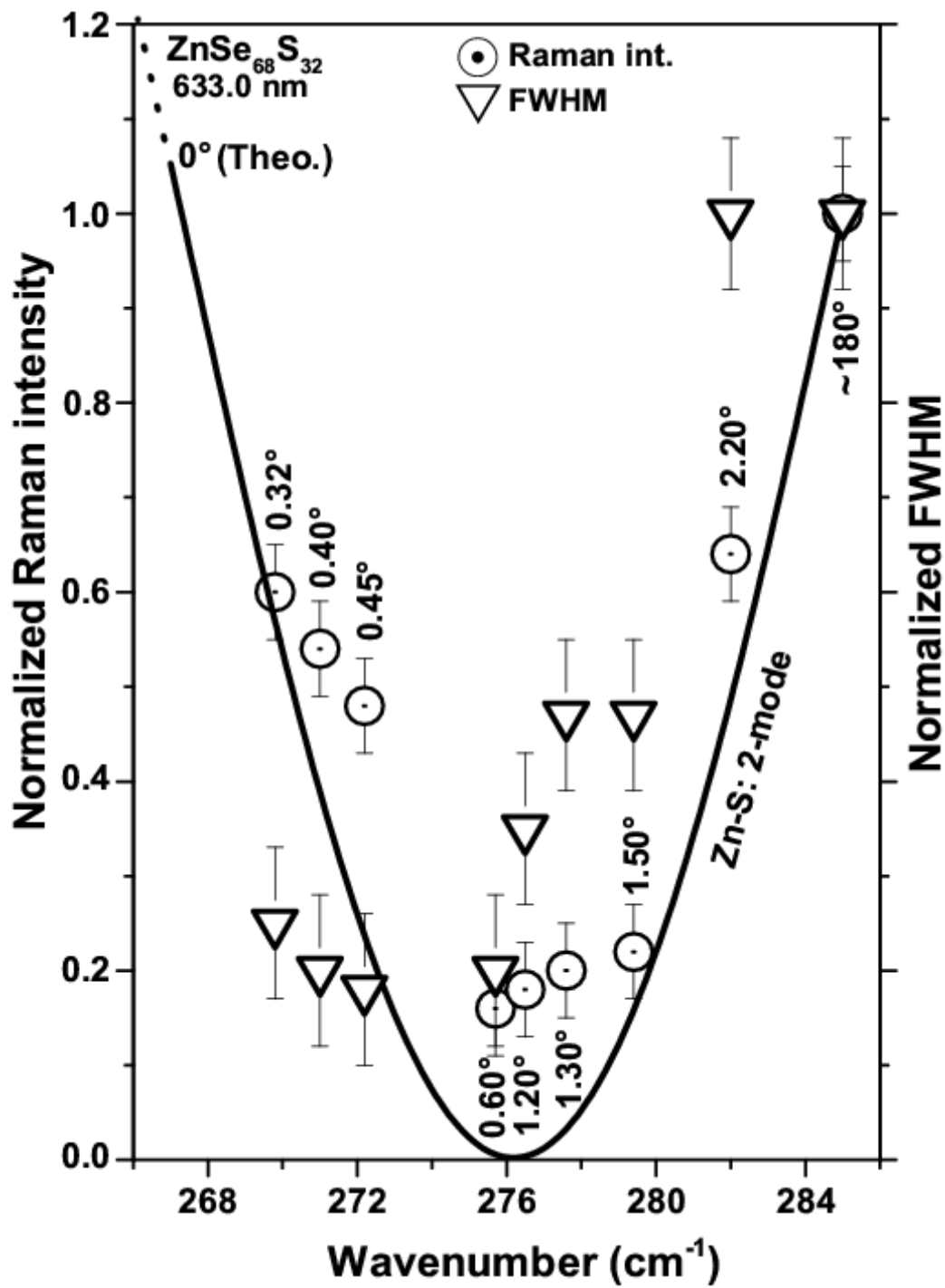

**Fig. A2**

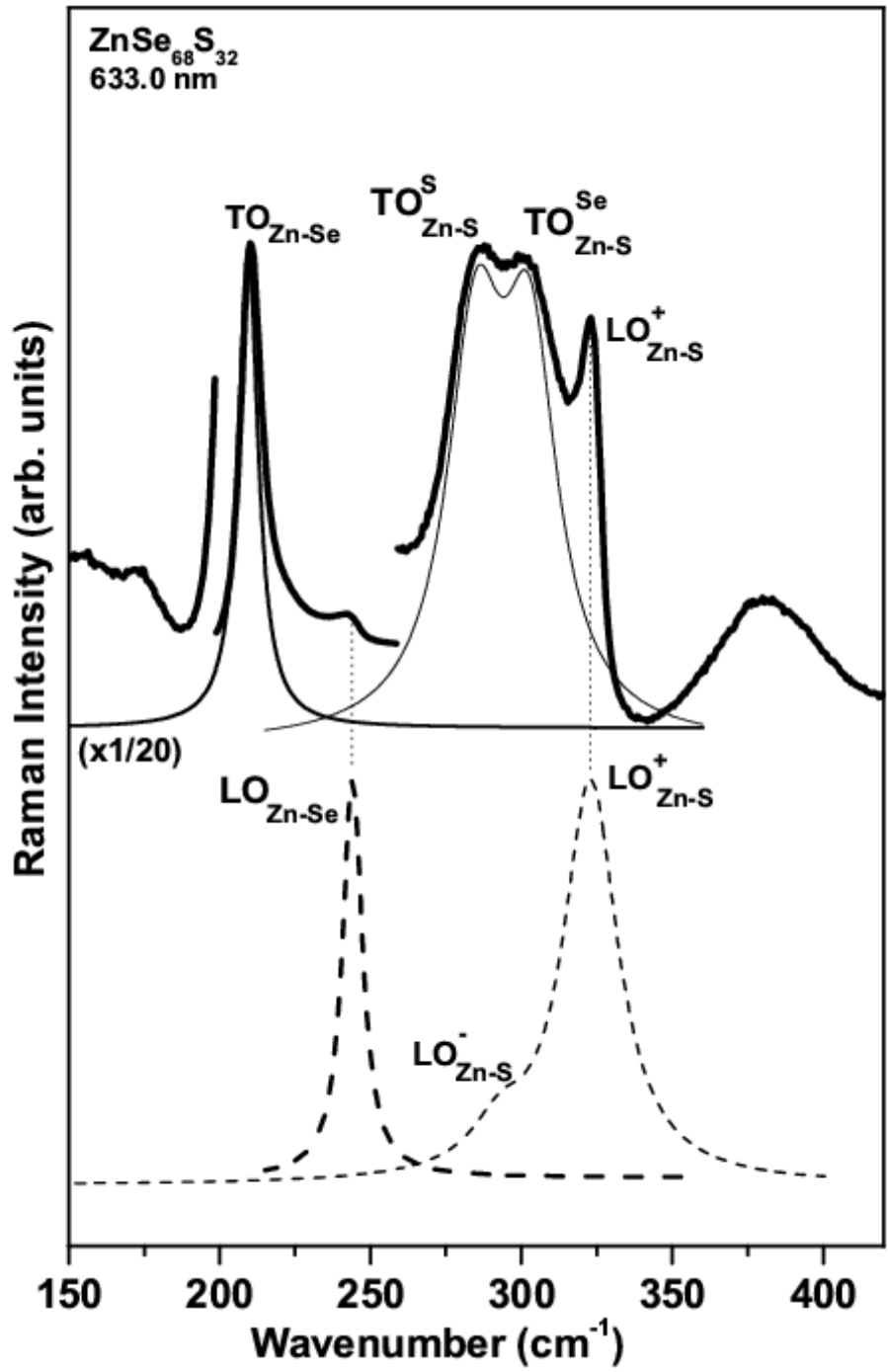

Fig. A3